\documentclass[proof]{WileyASNA-v1}

\newcommand{\angstrom}{\textup{\AA}}

\articletype{Article Type}%

\received{28 April 2019}
\revised{8 July 2019}
\accepted{8 July 2019}

\begin{document}

\title{Reverberation-mapping of distant quasars: time-lag determination using different methods}

\author[1]{M. Zaja\v{c}ek*}

\author[1]{B. Czerny}

\author[1]{M.-L. Martinez Aldama}

\author[2]{V. Karas}

\authormark{M. Zaja\v{c}ek \textsc{et al.}}

\address[1]{\orgdiv{Center for Theoretical Physics}, \orgname{Polish Academy of Sciences}, \orgaddress{ \country{Al. Lotnik\'ow 32/46, 02-668 Warsaw}, \state{Poland}}}

\address[2]{\orgdiv{Astronomical Institute}, \orgname{Academy of Sciences}, \orgaddress{ \country{Bo\v{c}n\'i II 1401, 141 00 Prague}, \state{Czechia}}}

\corres{*\email{zajacek@cft.edu.pl}}


\abstract{By applying different statistically robust methods, we analyze the time-lag between the continuum and ionized line-emission (Mg II line) light curves for the distant bright quasar CTS C$30.10$ (redshift $z \sim 0.9$). The data were obtained by the 10-meter SALT telescope in South Africa. In detail, we demonstrate the application of several methods using interpolated cross-correlation function (ICCF), discrete correlation function (DCF), z-transformed discrete correlation function (zDCF), von Neumann estimator, and the JAVELIN code package. In particular, we discuss the uncertainties of these methods. In conclusion, we find that the quasar lies on the broad-line region (BLR) size -- monochromatic luminosity power-law scaling, $R_{\rm BLR}\propto L_{5100}^{1/2}$, which was already confirmed for low-redshift sources. In case the BLR size-luminosity relation holds for other distant sources, quasars could be used for probing cosmological models as ``standard candles'' complementary to supernovae Ia.}


\jnlcitation{\cname{%
\author{M. Zaja\v{c}ek}, 
\author{B. Czerny}, 
\author{M.-L. Martinez Aldama}, and 
\author{V. Karas}} (\cyear{2019}), 
\ctitle{Reverberation-mapping of distant quasars: time-lag determination using different methods}, \cjournal{Astronomical Notes}, \cvol{2019; xxx}.}


\maketitle

\footnotetext{\textbf{Abbreviations:} AGN, active galactic nuclei; BLR, broad-line region; RM, reverberation mapping; MBH, massive black hole; FWHM, full width at half maximum; EW, equivalent width; VLTI, Very Large Telescope Interferometry}

\section{Introduction}

The reverberation mapping of quasars (hereafter RM) is a powerful technique that enables to measure the time-delay between the continuum (photometric) light curves in active galactic nuclei (hereafter AGN) and the reprocessed line-emission light curve. From the measured time-lag $\tau_0$, it is straightforward to estimate the lengthscale of the reprocessing region, $R\sim c\tau_0$ \citep[see e.g.][]{2000ApJ...533..631K,2004ApJ...613..682P,2018NatAs...2...63M}. It is thus an elegant method to shed light on the structure of the central region of active galaxies by determining the time-lags between the variable continuum and emission lines with a different ionization potential.

The reprocessing region is associated with the broad-line region (BLR), where the gaseous-dusty material is bound to the central massive black hole (hereafter MBH). The BLR is characterized by emission lines with the full width at half maximum (FWHM) of the order of $1000\,{\rm km\,s^{-1}}$ and more, with the characteristic values of $\Delta v_{\rm FWHM}\approx 5000\,{\rm km\,s^{-1}}$.

 The large width of the lines is consistent with the quasi-Keplerian motion of the gas in the gravitational potential of a single MBH, $\phi=-GM_{\bullet}/r$. The BLR then must be very compact, with the characteristic length-scale typically less than the innermost parsec, which directly follows from the Keplerian estimate of its size,

\begin{align}
\label{eq_blregion}
    r_{\rm BL,kin}&= f \frac{GM_{\bullet}}{v_{\rm K}^2}\\ 
                  &= 0.034f\left(\frac{M_{\bullet}}{2\times 10^8\,M_{\odot}}\right)\left(\frac{\Delta v_{\rm FWHM}}{5000\,{\rm km\,s^{-1}}}\right)^{-2}\,{\rm pc}\notag\\ 
                  &\sim 1\,800f\,R_{\rm s}\,\notag.     
\end{align}

Hence, the BLR has played a key role in the identification of AGN since it directly probes both the gravitational and the electromagnetic influence of the immediate surroundings of central MBHs, while the emitting gas is either bound to the MBH or experiences competing effects of the MBH gravitational potential and those of ionizing radiation field \citep{2011A&A...525L...8C}. In addition, the variability of the broad-line emission is strongly correlated with the variability of an ionizing continuum. The high correlation arises due to the reprocessing of the variable ultraviolet continuum photons from the central source by the BLR gas. The large equivalent widths of the broad lines imply a large covering factor of the central source, between $\sim 10-30\%$. On the other hand, Lyman continuum absorption lines due to BLR are difficult to detect \citep{1989ApJ...342...64A}. The combined requirement of the high covering factor and the absence of Lyman continuum absorption in most cases imply the flattened distribution of the BLR clouds -- that is a 	``nest geometry'' that we view through a hole for type I AGN \citep[see the review by][]{2009NewAR..53..140G}. This geometrical view is in accordance with the first kinematically resolved observation of the BLR in quasar 3C273 thanks to the Very Large Telescope Interferometry (VLTI) in near-infrared $K_{\rm s}$ band \citep[$2.2\,{\rm \mu m}$,][]{2018Natur.563..657G}. The corresponding spatial offset of the blue and the red photocenters of Pa$\alpha$ line is $0.03\,{\rm pc}$ along the plane perpendicular to the jet axis. In addition, the velocity profile is consistent with the Keplerian rotation of a flattened system of cloudlets and the spatial extent is in agreement within 2$\sigma$ uncertainties with the results obtained via the RM \citep{2019ApJ...876...49Z}.    

Another argument for a flattened system of the BLR clouds, which are dominantly in a Keplerian rotation, is the existence of the disk-like, double-peak Balmer line profiles that are approximately two times broader than single-peak Balmer lines \citep[$12\,500\,{\rm km\, s^{-1}}$ vs. $5\,700\,{\rm km\, s^{-1}}$][]{1994ApJS...90....1E,2004ApJS..150..181E}. The FWHM and line-profile dichotomy, which also correlates with the radio-quiet and radio-loud quasar (core/lobe-dominated) dichotomy \citep{1979ApJ...228L..55M}, seems to originate in the viewing angle \citep{1999ASPC..175..157G, 2004A&A...423..909P,2006NewAR..50..716B}, as the narrower single-peak emission lines are viewed nearly face-on, with inclinations less than $25^{\circ}$ as is supported by line fitting to both double-peak and single-peak sources \citep{1994ApJS...90....1E,2004ApJS..150..181E,2008SerAJ.177....9B}.  The line-fitting also showed that the turbulent velocity field perpendicular to the dominant Keplerian velocity field is needed to account for the broader line peaks. The turbulent velocity component is smaller than the Keplerian component, $v_{\rm Kepl}>v_{\rm turb}$, and was estimated to be $v_{\rm turb}\approx 1300\,{\rm km\,s^{-1}}$, with the 1-$\sigma$ scatter of $400\,{\rm km\,s^{-1}}$ \citep{1994ApJS...90....1E,2004ApJS..150..181E}.

The RM also revealed that higher-ionization lines have shorter time-lags and larger line-widths, and vice-versa for lower-ionization lines. Hence, this implies that the BLR has a stratified ionization structure, i.e. like a ``flattened onion'' with a hole along its axis \citep[as inferred from the RM-mapping campaign of NGC5548,][]{1991ApJ...366...64C,1991ApJ...368..119P,1993ApJ...408..416D,1993ApJ...404..576M}. The shorter time-delays and larger line widths for higher-ionization lines, which are presumably closer to the MBH, imply that the reprocessing, line-emitting gas is virialized, hence the BLR is a suitable probe of the gravitational influence of the black hole located at the very centre with the velocity inferred from the Doppler shift proportional to $r^{-1/2}$ \citep{1999ApJ...521L..95P,2000ApJ...540L..13P,2003A&A...407..461K,2010ApJ...716..993B}. This fact has enabled to determine virial black-hole masses beyond the local Universe, i.e. for distances larger than 150 Mpc, where spatially and kinematically resolved studies of stellar and gas motions suited for our Galaxy and other nearby galactic nuclei are no longer possible. In other words, a high spatial resolution is replaced by a high temporal resolution to measure the time-lag $\tau_0$ between AGN continuum variations, which arise from the inner portions of the accretion discs that emit thermal UV emission, and the corresponding broad emission-line response, which arises due to the gas photoionized by disc UV emission. In order to determine the time-delay, observations in the optical bands require the long-term monitoring campaigns lasting several months to years. 

Using the velocity of the BLR gas $v_{\rm BLR}$, which is estimated from the Doppler broadening of emission lines, i.e. from the observed line widths $\text{FWHM}_{\rm obs}$, it is possible to estimate the virial black hole mass assuming that the BLR gas is virialized,

\begin{equation}
   M_{\bullet}=\frac{R_{\rm BLR}v_{\rm BLR}^2}{G}=f_{\rm vir}\frac{c\tau_0 \text{FWHM}_{\rm obs}}{G}\,,
   \label{eq_virial_mass}
\end{equation} 
where $f_{\rm vir}$ is the virial factor, which is of the order if unity and depends on the geometry, kinematics and the emission properties of BLR clouds. In general, the virial factor depends on the line-of-sight inclination $i$ of the BLR plane (assuming the disc-like geometry of the BLR) as well as its thickness $H/R_{\rm BLR}$ \citep{2018NatAs...2...63M},

\begin{equation}
f_{\rm vir}=[4(\sin^2{i}+(H/R_{\rm BLR})^2)]^{-1}\,,
\label{eq_virial_factor}
\end{equation}
which introduces the overall uncertainty in the virial black-hole mass estimates. In fact, by fixing the virial factor to a constant value for all sources, as is often done for single-epoch measurements \citep{2015ApJ...801...38W}, the uncertainty range in the virial mass can differ by a factor of at least $2-3$ \citep{2018NatAs...2...63M}. \citet{2018NatAs...2...63M} compared the black hole masses inferred from the SED fitting of central accretion discs to virial masses and found that the virial factor is inversely proportional to the width of the observed broad lines, $f_{\rm vir}\propto \text{FWHM}_{\rm obs}^{-1}$. Such a dependence can arise from the line-of-sight inclinations of the BLR planar structure as well as radiation pressure effects on the BLR cloud distribution, which is consistent with the overall flattened distribution \citep{2009NewAR..53..140G,2018Natur.563..657G} that can be puffed up by radiation pressure \citep{2011A&A...525L...8C}, which leads to the overall nest-like geometry of the BLR that shares the plane of an accretion disc, at least for a low-ionization part of the BLR \citep[see the failed radiatively accelerated dusty outflow -- FRADO model,][]{2017ApJ...846..154C}.  

One of the results of the reverberation mapping is the radius-luminosity relation, $R_{\rm BLR}\propto L^{\alpha}$, which is the power-law relation between the BLR-size (or measured time-lags) and the optical luminosity of AGN, typically in the optical bands \citep{2006ApJ...644..133B,2009ApJ...697..160B,2013ApJ...767..149B}. The importance of the radius-luminosity relations lies in the fact that it can be utilized for estimating black-hole masses for galaxies across the cosmic history, just based on single-epoch spectroscopic measurements (by taking the measured optical luminosity of an AGN, from which one can estimate the BLR size and using the line-width, one obtains the black-hole mass). Because of the importance of the $R_{\rm BLR} - L$ relation, it is important to verify it for higher-redshift sources using the reverberation measurements. 

Another important aspect of the radius-luminosity relation is that it can be utilized to determine the luminosity of AGN and based on the measured flux, the luminosity distance can be obtained. Thus, the radius-luminosity relation provides a way to turn AGN into another population of standard candles that could be used for testing cosmological models over a larger range of redshifts than supernovae Ia \citep{2013ApJ...767..149B,2019arXiv190309687L}.

In this paper, we outline the way to reliably measure time-lags and thus BLR sizes for higher-redshift sources. We show this using the example of a bright quasar CTS C30.10 at $z=0.9$, for which the AGN continuum and MgII line-emission light curves were obtained. The detailed observational analysis was shown in \citet{2014A&A...570A..53M} and \citet{2019arXiv190109757C}. Here we provide an overview of the time-lag determination methods, which provide statistically robust results even for sparse and heterogenuous time series of more distant sources.

The paper is structured as follows. In Section~\ref{sec_observations}, we provide an overview of observational data of CTS C30.10. Subsequently, in Section~\ref{sec_time_lags}, an overview of time-lag determination methods is provided, including interpolated cross-correlation function, discrete correlation fuction, $z$-tranformed discrete correlation function, JAVELIN algorithm, and the optimized von Neumann scheme. Finally, we summarize the main results in Section~\ref{sec_summary}.

\section{Observational Data}\label{sec_observations}

The studied source CTS C30.10 was found in the Calan-Tololo Survey as a bright quasar \citep{1993RMxAA..25...51M} of $V=17.2$ (NED magnitude) at the intermediate redshift of $z=0.90052$ \citep{2014A&A...570A..53M}. The source is located in the southern hemisphere with the equatorial coordinates of RA = 04h47m19.9s, Dec=-45d37m38s (J2000.0). Between December 6, 2012 and December 10,2018 it was monitored by the  10-m SALT telescope in order to reliably detected the time-lag of MgII line with respect to the continuum, determine the size of the BLR, and finally to probe the radius-luminosity relation towards higher redshifts.

The MgII emission-line light curve is constructed based on a long-slit spectroscopy using the SALT telescope in a service mode, using the slit width of 2''; for further details please read \citet{2014A&A...570A..53M} and \citet{2019arXiv190109757C}. The MgII line information was extracted from the quasar spectra in the wavelength range between $2700$ and $2900\,\angstrom$ by fitting different spectral components to the spectra, namely the power-law component corresponding to the accretion disc emission, FeII-line pseudo-continuum, and the two kinematic components of MgII line (redshifted and blueshifted Lorentzians), each of which furthermore consists of a doublet at $2796.35\, \angstrom$ and $2803.53\, \angstrom$, with the doublet ratio of $1.6$. The equivalength width (EW) of MgII was calculated by integrating all four spectral components (2 kinematics plus 2 doublets) in the band $2700$-$2900\,\angstrom$.

The spectroscopy information was complemented by the photometric lightcurve \citep{2014A&A...570A..53M, 2019arXiv190109757C}, which is generally denser than the MgII-line light curve. The photometric observations of CTS C30.10 were performed using OGLE-IV survey ($V$-band) and SALTICAM (g-band). In the construction of the final continuum light curve used in this paper, the SALTICAM measurements were allowed to vary by an arbitrary shift to match the overlapping $V$-band flux densities. The MgII-line flux density was inferred using the continuum light curve in the $V$-band, since the SALT observations are not spectrophotometric due to the construction of the telescope\footnote{For more information, please visit, \url{https://www.salt.ac.za/}}. The continuum flux density was corrected for the dust extinction as taken from NED. The $V$-band values were then used to calibrate the power-law component in the spectra and MgII-line flux was determined from the fitted kinematic components using the numerically calculated EW.

The total normalized continuum dispersion was $6.0\%$ and the line dispersion was $5.2\%$, a bit lower than the continuum variability amplitude, but comparable within uncertainties. The continuum and MgII line-emission light curves were then utilized for the subsequent time-lag analysis. 

\section{Determination of time-lags}\label{sec_time_lags}

In this section, we list several statistically robust methods that are suitable for analyzing heterogeneous and generally uneven and sparse astronomical time series.

\subsection{Interpolation Cross-Correlation Function}

A traditional method to evaluate how much (anti)correlated two time series are is to calculate the cross-correlation function defined as,

\begin{equation}
  r=\frac{\sum_{i=1}^{N}(x_i-\overline{x})(y_i-\overline{y})}{\sqrt{\sum_{i=1}^N (x_i-\overline{x})^2}\sqrt{\sum_{i=1}^N (y_i-\overline{y})^2}}\,,
  \label{eq_correlation_coefficient}
\end{equation}
where $(x_i,y_i)$ are N points of the time series. For astronomical measurements, the two time series are not evenly sampled and the usual procedure is to linearly interpolate one series with respect to the other one. In the context of reverberation mapping, the cross-correlation function CCF($\tau$) is calculated first for the continuum points interpolated to match time-shifted spectroscopic points to get $\text{CCF}_1({\tau})$, and then the spectroscopic points are interpolated to match time-shifted photometric points to obtain $\text{CCF}_2({\tau})$. The final cross-correlation function value is simply obtained by the average, $\text{CCF}({\tau})=0.5[\text{CCF}_1(\tau)+\text{CCF}_2(\tau)]$. 

For calculating the centroid and the peak of the CCF, we use the algorithm implementation of \citet{1998PASP..110..660P}; see also the software application as presented by \citet{2018ascl.soft05032S}. The algorithm cross-correlates the continuum and MgII line-emission light curves and determines the centroid and the peak of the cross-correlation function. Finally, we also run 5000 Monte Carlo simulations to determine the centroid and peak distributions and their corresponding uncertainties using the flux randomization technique. The centroid and peak values for the observer frame inferred from the CCF analysis are summarized in Table~\ref{table_CCF} for the symmetric interpolation, continuum-only, and line-only interpolations. 

\begin{table}[h!]
  \centering
  \caption{CCF centroid and peak values and their corresponding uncertainties for the symmetric interpolation as well as continuum-only and line-only interpolations.}
  \resizebox{0.5\textwidth}{!}{  
  \begin{tabular}{c|c|c}
     \hline
     \hline
     Interpolation mode & CCF Centroid [days] & CCF Peak (days)\\
     \hline
     Both & {\bf $1033.7^{+31.1}_{-554.7}$} & $1035.0^{+35.0}_{-550.0}$\\
     Continuum & $475.2^{+536.8}_{-187.2}$ & $480.0^{+555.0}_{-175.0}$\\
     Line & $1050.0^{+14.8}_{-447.0}$ & $1050.0^{+5.0}_{-447.0}$\\
     \hline
  \end{tabular}
  }
  \label{table_CCF}
\end{table}   

\begin{figure*}[h!]
  \centering
  \includegraphics[width=\textwidth]{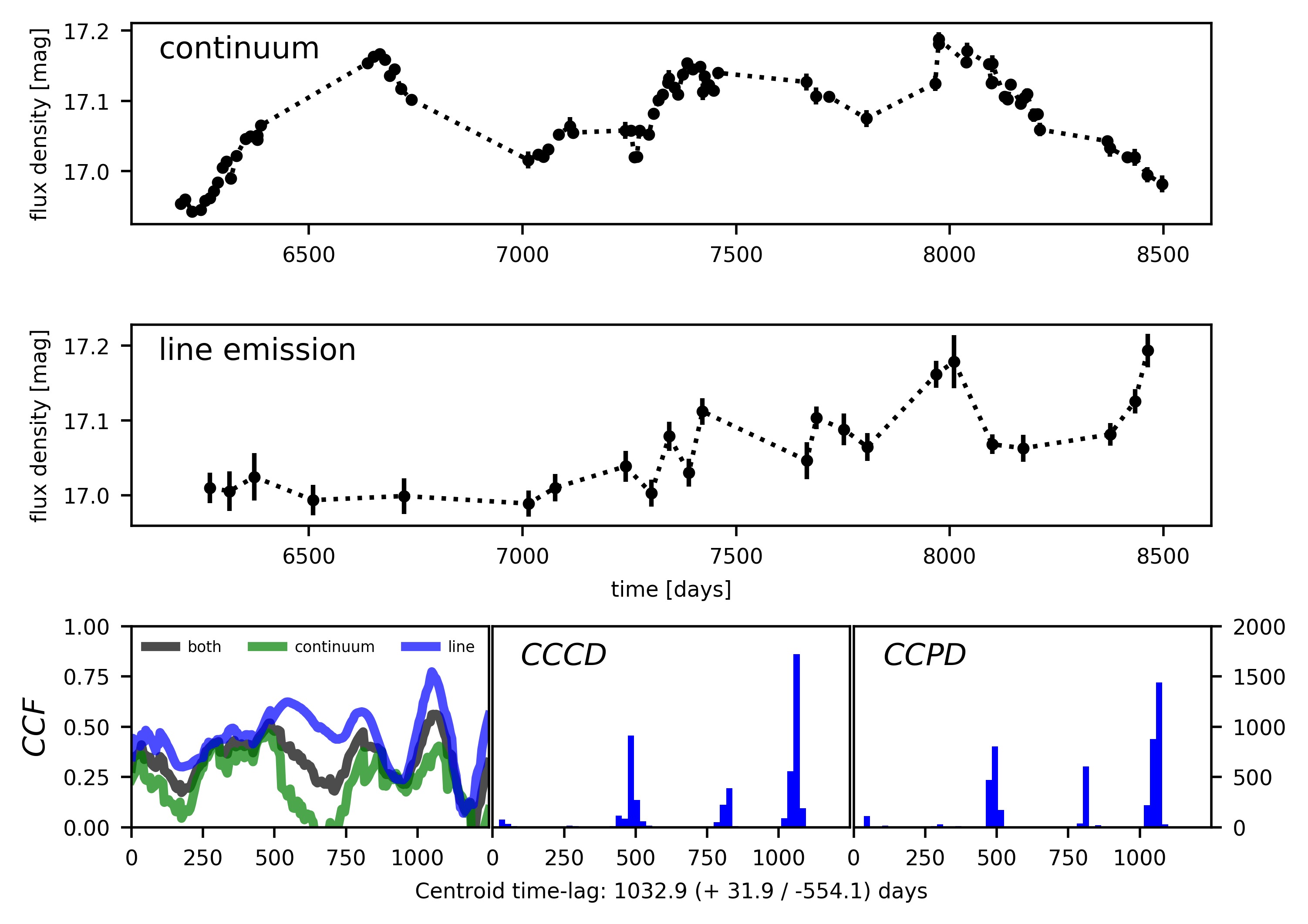}
  \caption{Cross-correlation analysis of the continuum and line-emission light curves (upper two panels). The bottom panels show the fundamental results of the CCF analysis: left panel--CCF for symmetric interpolation (black line) as well as the continuum-only interpolation (green line) and the line-emission-only interpolation (blue line); middle panel -- CCF centroid distribution; right panel -- CCF peak distribution. The time-lags are expressed in the observer frame.}
  \label{img_CCF}
\end{figure*}

In Fig.~\ref{img_CCF}, we show the used continuum and line-emission data as well as the CCF distribution (for interpolating both light curves -- black line, interpolated continuum -- green line, interpolated line-emission -- blue line). The results of the Monte Carlo flux-randomization realizations are depicted in the bottom middle and right panels that show the cross-correlation centroid and peak distributions for the symmetric interpolation case. In the further analysis, we consider the CCF centroid value for the symmetric interpolation case, $\tau_0=1033.7^{+31.1}_{-554.7}$ days.

\subsection{Discrete correlation function}

\begin{figure*}[h!]
  \centering
  \begin{tabular}{cc}
    \includegraphics[width=0.5\textwidth]{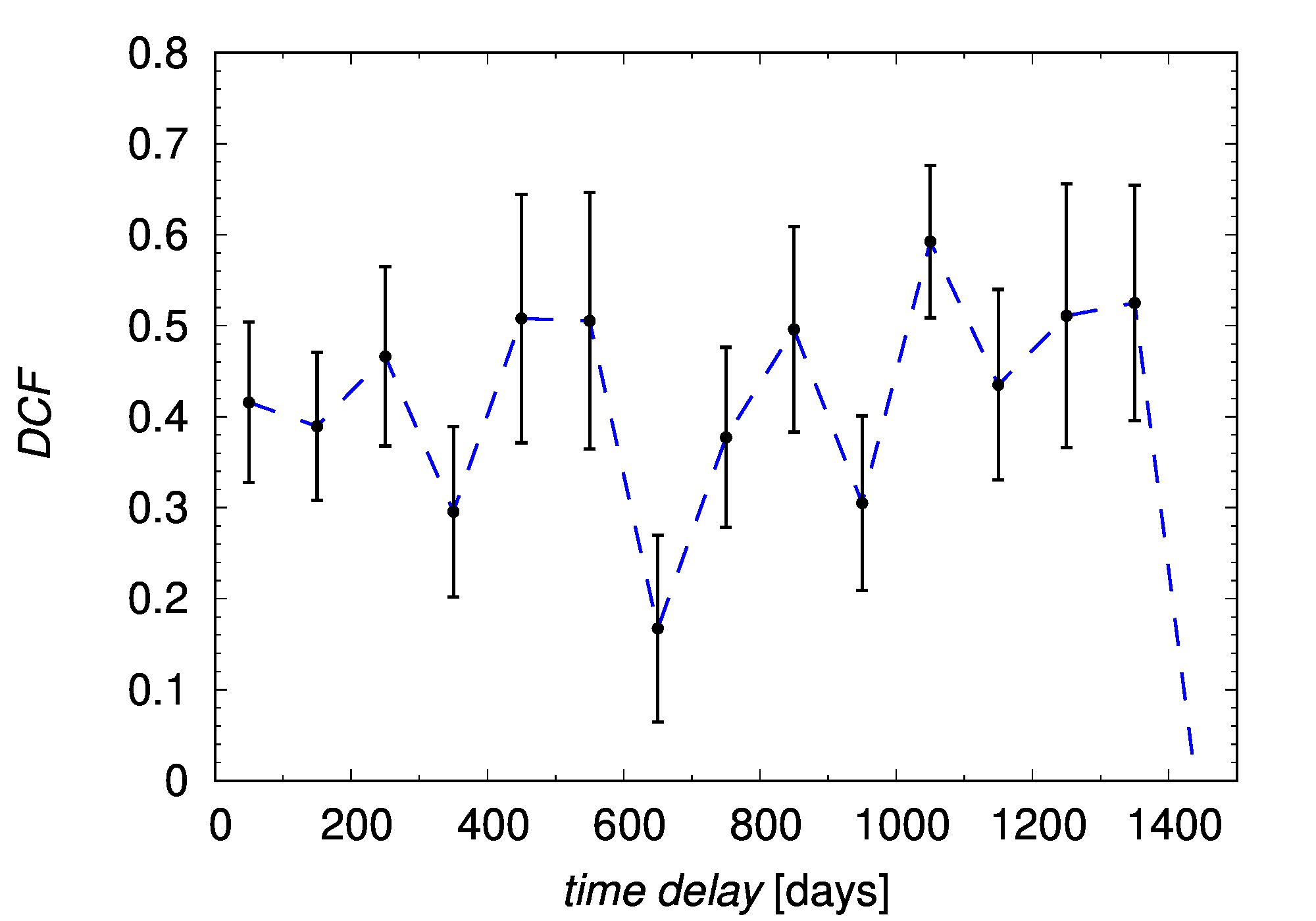} & \includegraphics[width=0.5\textwidth]{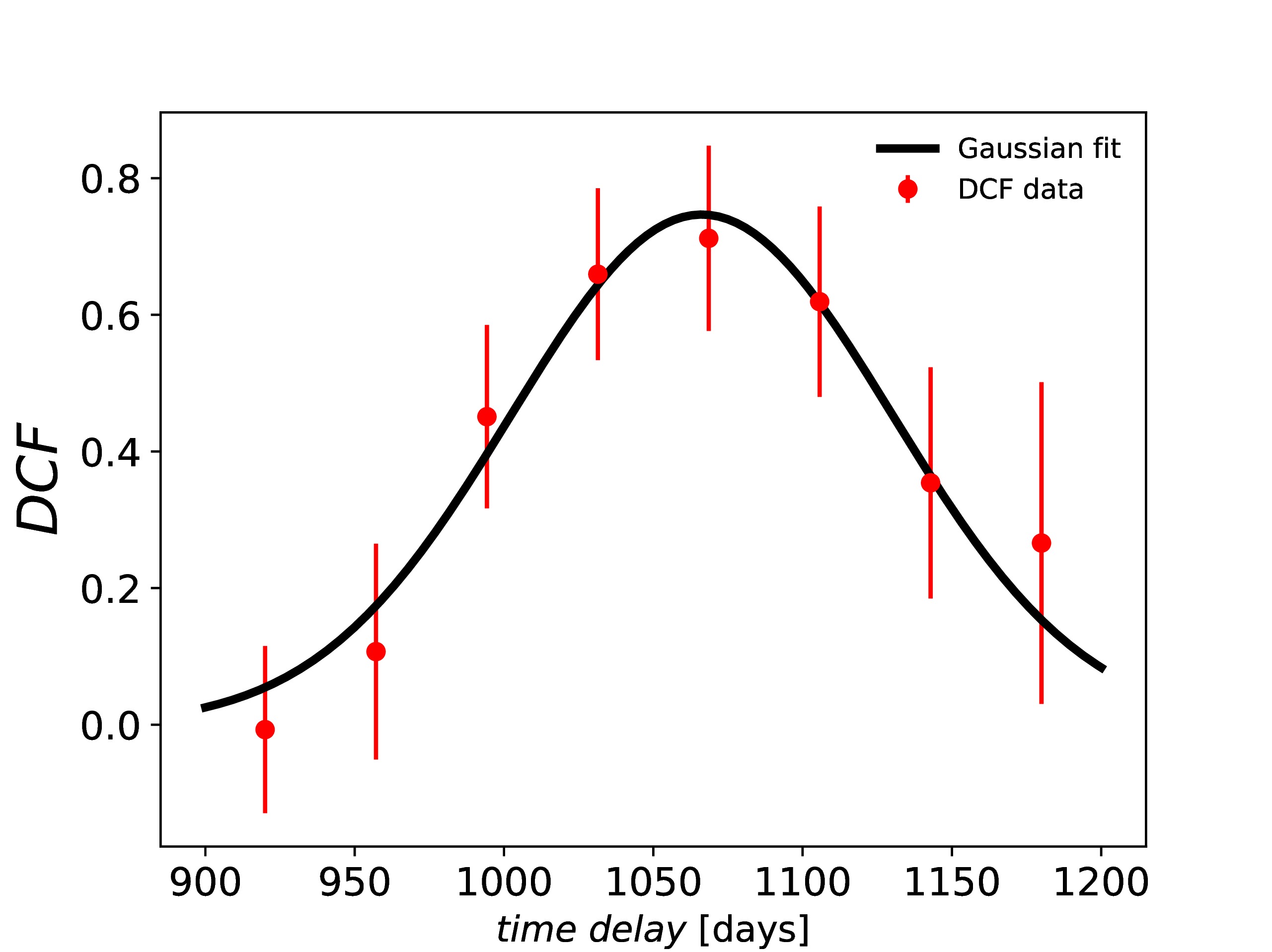}
  \end{tabular}
  \caption{\textbf{Left panel:} DCF values in the time-lag interval of 0-1500 days, with the bin size of 100 days. \textbf{Right panel:} The region close to the dominant peak of 1050 days. We fitted the Gaussian to the peak surroundings to obtain the mean peak value and its uncertainty, $\tau_0=1066\pm 64$ days. The time-lags are expressed in the observer frame.}
  \label{fig_dcf}
\end{figure*}

In comparison with the interpolated CCF, the advantage of a discrete correlation function (DCF) is that it does not introduce additional points into the time series, which can effect the overall variability of light curves. In the context of time-lag determination, it was suggested by \citet{1988ApJ...333..646E} with the focus on unevenly sampled data. The DCF is defined based on the unbinned CCF (UCCF) between each measured pair of the light curves $(x_i,y_j)$,

\begin{equation}
  \text{UCCF}_{ij}=\frac{(x_i-\overline{x})(y_j-\overline{y})}{\sigma_x\sigma_y}\,.
  \label{eq_uccf}
\end{equation}
DCF is then $\text{UCCF}_{\rm ij}$ binned into equal time bins. Given our photometric and spectroscopic light curves, the mean time between continuum observations is $\overline{\Delta t}_{\rm cont}=T/n=(8497.29000-6199.79884)/82 \approx 28$ days and the corresponding mean time for the MgII line curve is  $\overline{\Delta t}_{\rm line}=T/n=(8463.50000-6268.50000)/23 \approx 95$ days. Therefore, we choose the time-step at least as large as the larger of the two mean times, i.e. $\Delta t=100$. We note that the DCF values depend quite strongly on the chosen time-step, since the population size per equal bins varies considerably. The largest value of the DCF is for the time-lag bin of $\tau_0=1050$ ($\text{DCF}=0.59$), see the left panel of Fig.~\ref{fig_dcf}. Consequently, we focus on the time-lag region close to $\tau_0=1050$ peak. In the range of time-lag bins $\tau=900-1200$ days, we run the DCF analysis with the time-bin of 40 days. Then, we fit the DCF values with the Gaussian to obtain the mean value and its uncertainty of $\tau_0=1066\pm 64$ days, see Fig.~\ref{fig_dcf} (right panel).

\subsection{Z-Transformed Discrete Correlation Function}

\begin{figure}[h!]
  \centering
  \includegraphics[width=0.5\textwidth]{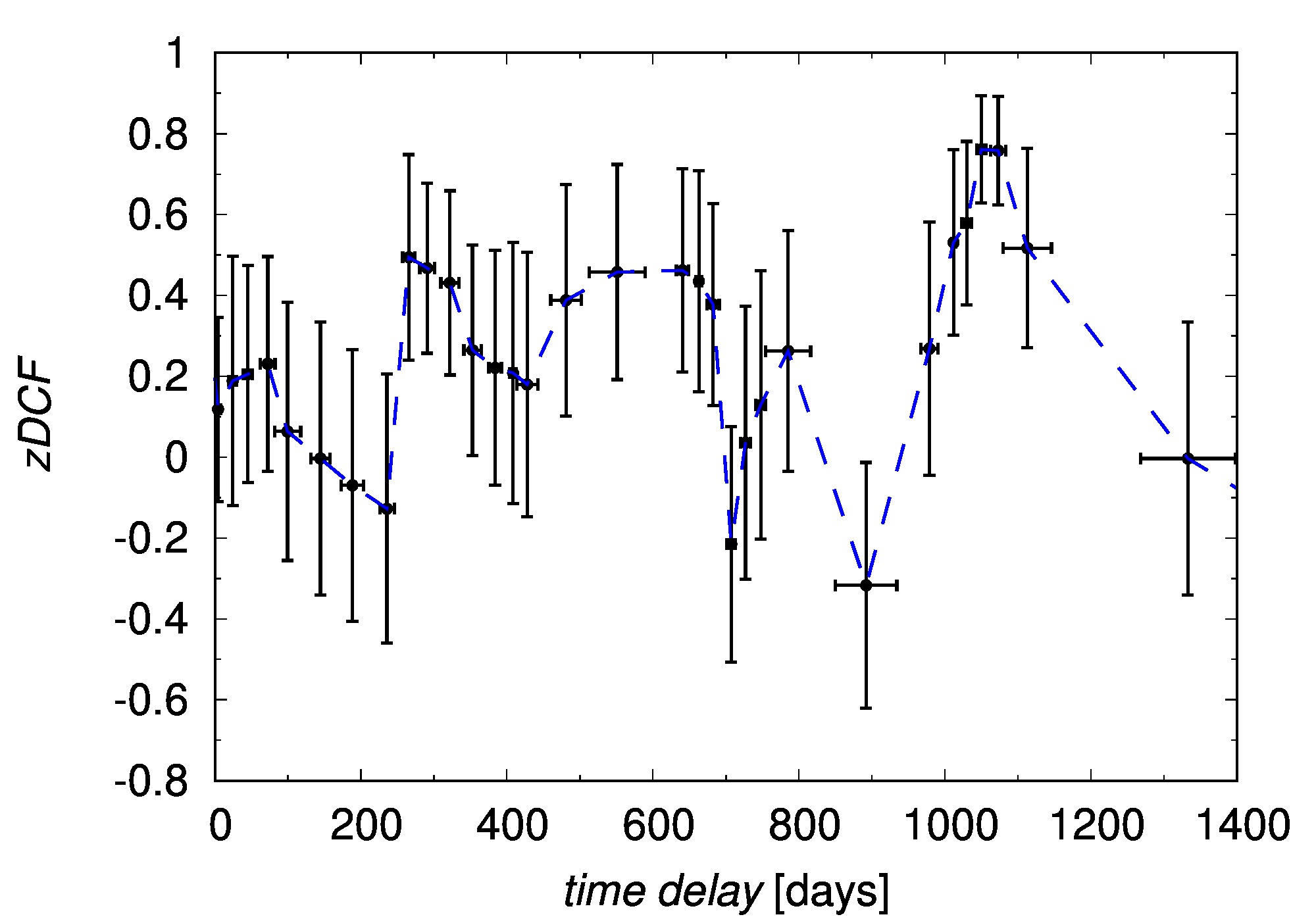}
  \caption{zDCF values as a function of the time-delay (in days) in the observer frame, including the uncertainties.}
  \label{img_zdcf}
\end{figure}

Because of several biases, DCF was corrected by its $z$-transformation using Fisher's $z$-transform as well as using equal population binning rather than equidistant time-step \citep{1997ASSL..218..163A}. The $z$-transformed DCF (zDCF) is therefore better suited for undersampled, sparse and heterogeneous pairs of light curves than classical DCF\footnote{For the numercal algorithm related to zDCF and its uncertainties, please visit \url{http://www.weizmann.ac.il/particle/tal/research-activities/software}}. In fact, the light curve can have as few as 12 points and zDCF does not assume neither AGN variability model not smoothness of light curves. In addition, the uncertainties can be calculated using Monte Carlo averaged zDCF values as inferred from generated pairs of light curves with random errors. In Fig.~\ref{img_zdcf}, we plot zDCF values with the corresponding uncertainties versus the time-lag value. As in previous time-lag analyses, several peaks are apparent, with the dominant one at $\tau_0\sim 1050$ days with the zDCF value of $\text{zDCF}\sim 0.76$. To make sure that this peak is most likely, we applied the maximum likelihood function of \citet{1997ASSL..218..163A} in the interval between 800 and 1400 days and we obtained the peak as well as its uncertainties, $\tau_0=1050^{+54}_{-27}$ days. 

\subsection{Time-lag determination using the JAVELIN code}

Taking into account the AGN continuum variability, which can be modelled as a stochastic process and specifically as a damped random walk \citep[DRW;][]{2009ApJ...698..895K,2010ApJ...721.1014M,2010ApJ...708..927K,2016ApJ...826..118K}, the line-emission light curve can be treated as a scaled, smoothed, and time-lagged response to the continuum emission. Based on these principles, JAVELIN (Just Another Vehicle for Estimating Lags In Nuclei) code was developed \citep{2011ApJ...735...80Z,2013ApJ...765..106Z,2016ApJ...819..122Z}\footnote{For more information, please visit \url{https://bitbucket.org/nye17/javelin/src/develop/}}. The JAVELIN code makes use of the Monte Carlo Markov Chain (MCMC) to infer posterior probabilities of two AGN-continuum variability parameters (characteristic timescale and variability amplitude), and the probability distributions of the emission-line time-lag, the smoothing width of the top-hat function, as well as the scaling ratio between the line and the continuum amplitudes ($A_{\rm line}/A_{\rm cont}$).

\begin{figure}[h!]
   \centering
   \includegraphics[width=0.5\textwidth]{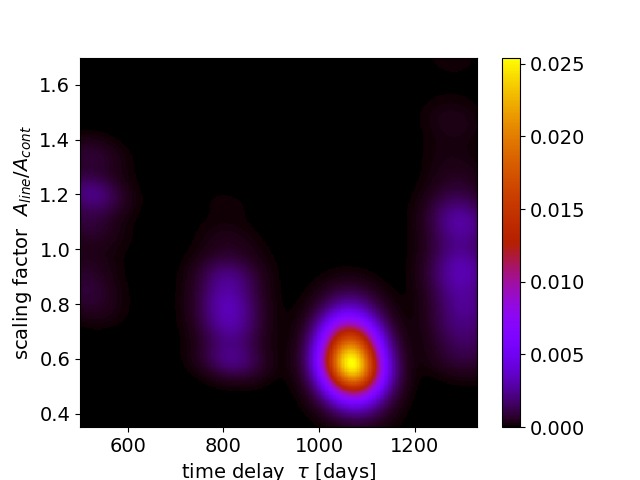}
   \caption{JAVELIN MCMC results in the time lag--scaling factor plane. The dominant peak is clearly visible at $\tau=1068$ days. The time lag is expressed with respect to the observer frame.}
   \label{img_javelin}
\end{figure}

In Fig.~\ref{img_javelin}, we plot the JAVELIN posterior probability results in the time lag (observer frame) -- scaling ratio ($A_{\rm line}/A_{\rm cont}$) plane. As in previous time-delay measurement methods, several peaks are apparent, but the strongest one is at $\tau_0=1068^{+211}_{-252}$ days. The uncertainty was determined based on the total distribution of time-lags, taking into account the surrounding minor peaks.

\subsection{Measures of the regularity of data -- von Neumann estimator}

Previous time-lag analyses methods are complemented by a rather novel technique in the astronomical time-series analysis, which measures the randomness (or regularity) of data. Such a technique is used in cryptography and electronic data-compression methods. The main advantage is that estimators of the data randomness do not require any interpolation, such as for the interpolated CCF, and they do not need binning in the correlation space as for DCF or zDCF. In addition, no assumption concerning the AGN variability is needed as for JAVELIN code, hence this method is model-independent. Data randomness or complexity may be evaluated by various estimators, among which the optimized von-Neumann estimator appears to be most robust \citep{2017ApJ...844..146C}. The von Neumann estimator is defined based on the mean successive difference of the combined lightcurve $F(t,\tau)=\{(t_i,f_i)\}_{i=1}^{N}=F_1 \cup F_2^{\tau}$, 

\begin{equation}
  E(\tau)\equiv \frac{1}{N-1}\sum_{i=1}^{N-1}[F(t_i)-F(t_{i+1})]^2\,,
  \label{eq_von_Neumann}
\end{equation}  
where $F(t,\tau)$ is the combination of the continuum and time-lagged line-emission light curves. When the time-lag is close to the actual time-lag, $\tau\sim \tau_0$, the von Neumann estimator $E(\tau\sim \tau_0)$ reaches the minimum.

\begin{figure*}[h!]
  \centering
  \begin{tabular}{cc}
    \includegraphics[width=0.5\textwidth]{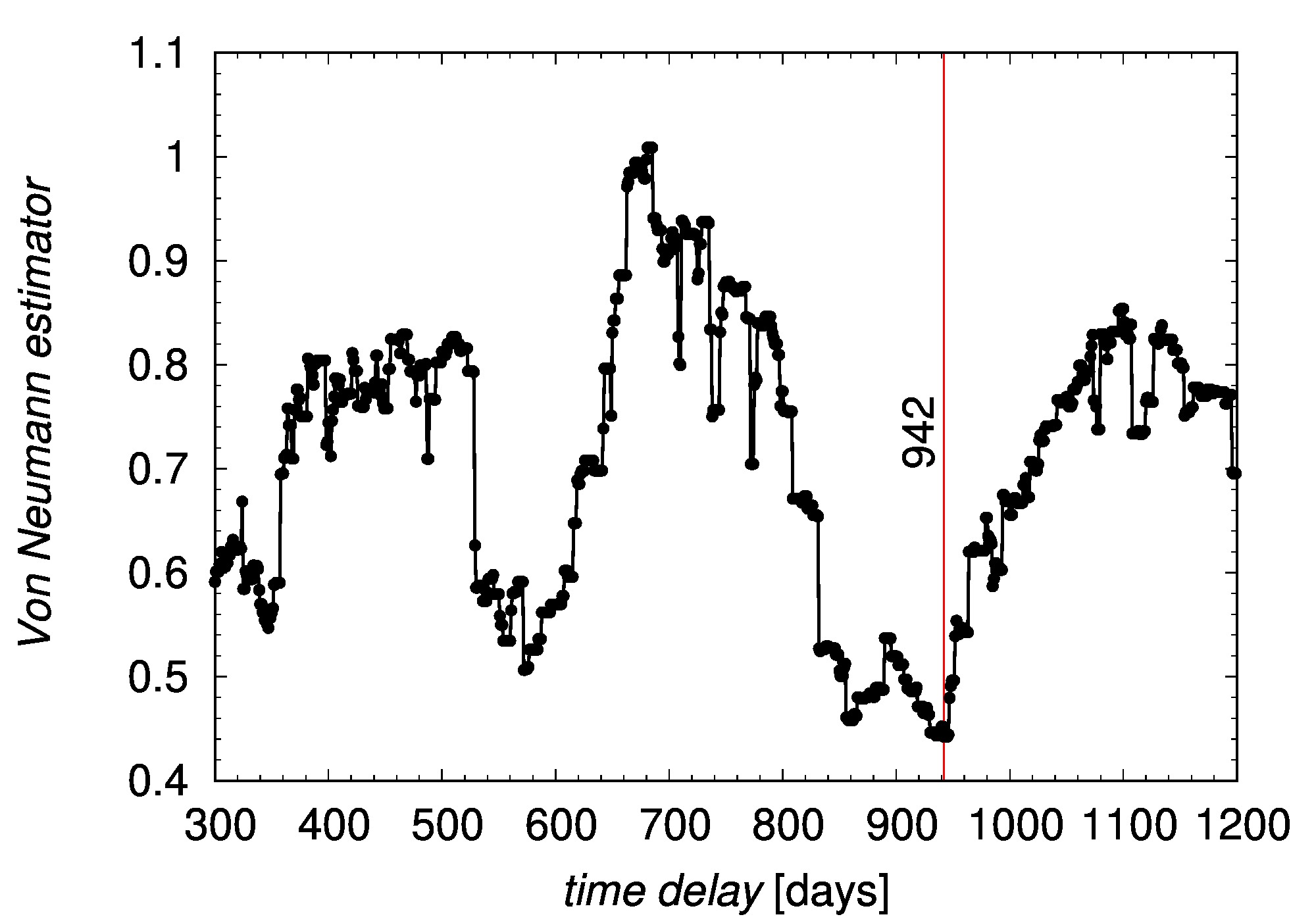} & \includegraphics[width=0.5\textwidth]{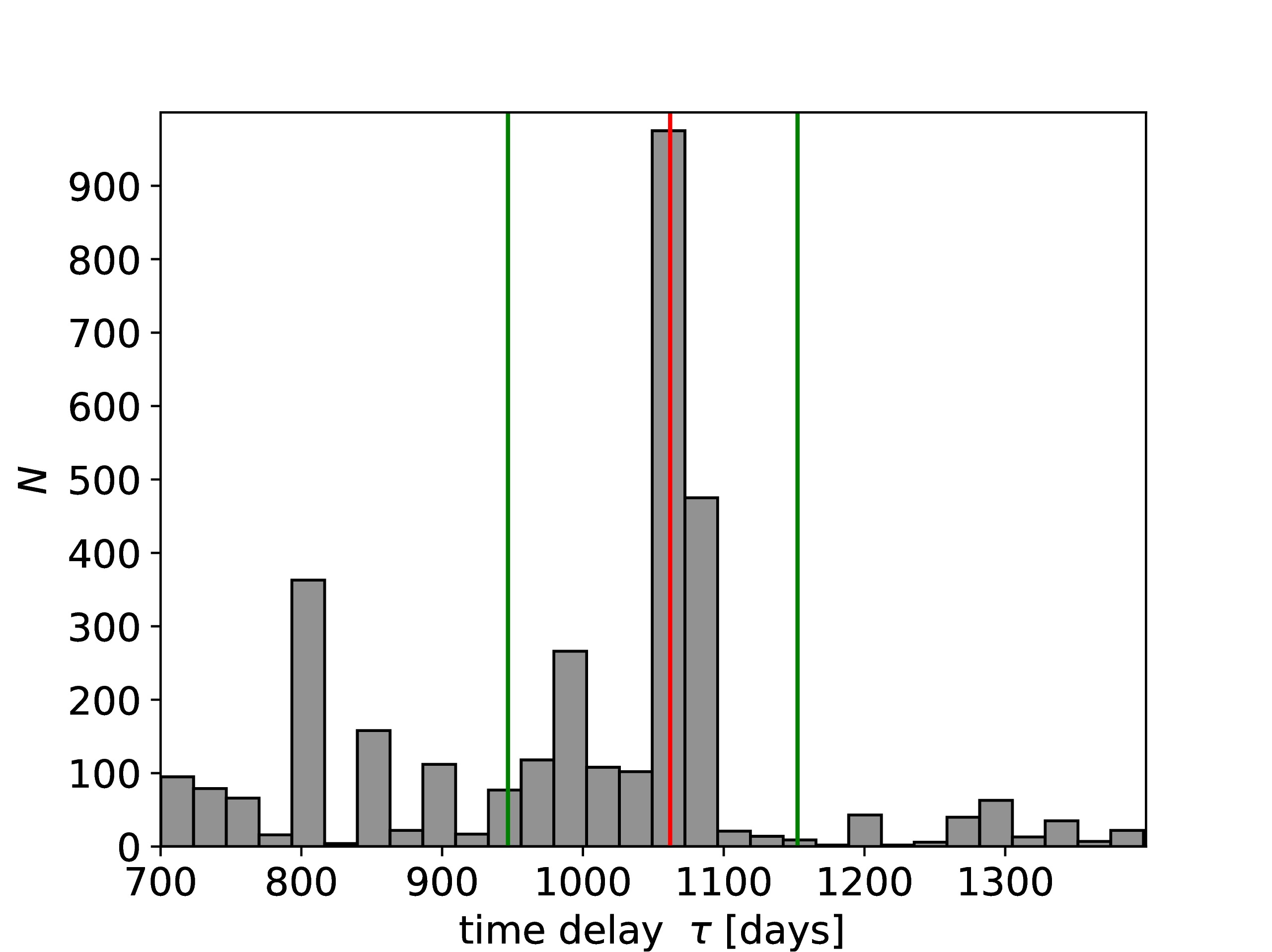}     
  \end{tabular}
   \caption{\textbf{Left panel:} Von Neumann estimator values with respect to the time-lag in the observer frame. The minimum value at $\tau\sim 942$ days is depicted by a vertical line. \textbf{Right panel:} The distribution of the minima from the optimized von-Neumann-estimator analysis as a result of 3000 bootstrap realizations. The mean value with the 1$\sigma$ errors is shown by vertical lines, $\tau_0=1062.0^{+90.5}_{-115.2}$ days.}  
  \label{img_von_neumann}
\end{figure*}

For CTS C30.10 source, we applied von Neumann estimator as expressed in Eq.~\eqref{eq_von_Neumann} using the \texttt{python} script of  \citep{2017ApJ...844..146C}\footnote{See also more at \url{www.pozonunez.de/astro_codes/python/vnrm.py}.}. We modified the code to add the calculation of uncertainties using the \textit{bootstrap} method, i.e. by generating pairs of light curves with the omission of at least $1/e=36.8\%$ of the original points (for the classical bootstrap, subsamples of original samples are created with a certain fraction of the points replaced by other points in the sample. For the time series, the identical points are then omitted, which leads to the further undersampling of the data). In Fig.~\ref{img_von_neumann}, we plot the von-Neumann estimator values with respect to the time-lag in the observer frame (left panel) with the minimum value at $\tau\sim 942$ days. To verify statistically the distribution of the minima, we performed 3000 bootstrap realizations. The final distribution is in the right panel of Fig.~\ref{img_von_neumann}, where we show the maximum value with the associated 1$\sigma$ errors, $\tau_0=1062.0^{+90.5}_{-115.2}$ days in the observer frame.  

\section{Summary}\label{sec_summary}

We summarize the time-lag results in both the observer and co-moving frame in Table~\ref{table_timelag} for the interpolated CCF, DCF, zDCF, JAVELIN, and von Neumann scheme. In general, the results are consistent with the mean value of the time-lag in the observer frame of $\tau_0=1056$ days. The difference lies in the uncertainties, which are significantly larger mainly for interpolated CCF and JAVELIN schemes. Generally, the uncertainties vary mainly due to different methods that are used to estimate them. In case of the interpolated CCF and JAVELIN, other time-lag peaks were considered while estimating the error, which led to its larger value in comparison with DCF, zDCF, and von Neumann scheme, for which the focus was on the surroundings of the main peak. The average value of the time-lag has an asymmetric error mainly due to the fact that there are more secondary peaks shortward of the dominant peak than in the opposite, longward direction, which approaches the total duration of SALT observations.  

\begin{table}[h!]
\caption{Overview of the applied time-lag determination methods for the source CTS C30.10 and the corresponding time-lags with uncertainties. The bottom part contains the final average values of the time lag in both the observer and the comoving frames.}
  \begin{tabular}{c|c}
    \hline
    \hline
    Method & Time-lag [days]\\
    \hline
    Interpolated CCF & $1034^{+31}_{-555}$\\
    DCF & $1066\pm 64$\\
    zDCF & $1050^{+54}_{-27}$\\
    JAVELIN & $1068^{+211}_{-252}$\\
    Von Neumann & $1062^{+91}_{-115}$\\
    \hline
    Average (observer frame) & $1056^{+51}_{-125}$\\
    Average (comoving frame) & $556^{+27}_{-66}$\\
    \hline
  \end{tabular}
  \label{table_timelag}
\end{table}

The average value in the comoving frame $\overline{\tau}_{\rm comov}=556^{+27}_{-66}$ days corresponds to the physical length-scale of the BLR, $R_{\rm BLR}=c\overline{\tau}_{\rm comov}=0.47^{+0.02}_{-0.06}\,{\rm pc}$, which is consistent with typical subparsec values, see also Eq.~\eqref{eq_blregion} and the direct detection of the BLR by \citet{2018Natur.563..657G}. 

In addition, the rest-frame time-lag of $\overline{\tau}_{\rm comov}=556^{+27}_{-66}$ days is consistent within uncertainties with the radius-luminosity relationship \citep{2019arXiv190109757C}, which was previously studied and determined for a lower-redshift sample of quasars \citep{2013ApJ...767..149B}. In principle, this opens up a possibility of using quasars for constraining cosmological parameters thanks to the validity of radius-luminosity relationship towards higher redshifts \citep{2019arXiv190309687L}. In parallel, quasars have been employed in cosmological studies using the X-ray and UV-luminosity relationship and its application to the Hubble diagram of quasars \citep{2019NatAs.tmp..195R}. Furthermore, with the combination of the spectroastrometric measurements using GRAVITY VLTI instrument \citep{2018Natur.563..657G} and the RM, it is possible to determine cosmological distances and constrain the Hubble constant \citep{2019arXiv190608417W}. In summary, quasars will play a crucial role in testing cosmological models, mainly towards redshifts larger than one, complementing supernovae Ia measurements.


\section*{Acknowledgments}

We acknowledge the financial support of the \fundingAgency{National Science Centre}, Poland, grant No. \fundingNumber{2017/26/A/ST9/00756} (Maestro 9) as well as the financial contribution of \fundingAgency{Czech Science Foundation - Deutsche Forschungsgemeinschaft collaboration} 
project No. \fundingNumber{19-01137J}.

\subsection*{Author contributions}

M. Zaja\v{c}ek put together the text, basic analysis, and the plots. B. Czerny, M.-L. Martinez Aldama and V. Karas provided additional analysis as well as comments.

\subsection*{Financial disclosure}

National Science Centre, Poland, grant No. 2017/26/A/ST9/00756 (Maestro 9).\\
Czech Science Foundation - Deutsche Forschungsgemeinschaft collaboration 
project No. 19-01137J

\subsection*{Conflict of interest}

The authors declare no potential conflict of interests.





\bibliographystyle{Wiley-ASNA}%
\bibliography{zajacek}

\section*{Author Biography}

\begin{biography}{\includegraphics[width=60pt,height=70pt]{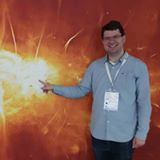}}{\textbf{Michal Zaja\v{c}ek.} Michal Zaja\v{c}ek, PhD., finished bachelor studies in general physics at the Charles University in
Prague in 2012, after defending the bachelor thesis ``Late Heavy Bombardment at various places of the Solar System" supervised by Miroslav Bro\v{z}, PhD. Subsequently, he continued with the master degree in Astronomy and Astrophysics with the thesis ``Neutron stars near the galactic centre", supervised by Prof. Vladim\'{i}r Karas, which he defended at the
same university in 2014. In the autumn of 2014, he started doctoral studies at the International Max Planck Research School based at the University of Cologne, Germany,
and Max Planck Institute for Radioastronomy in Bonn. In October 2017, he
defended the PhD Thesis ``Interaction between interstellar
medium and black hole environment", which was supervised
by Prof. Dr. Andreas Eckart and Prof. Dr. Anton J. Zensus. Between October 2017 and January 2019, he was a postdoctoral fellow in the VLBI group (Prof. Dr. Anton J. Zensus) at the
Max Planck Institute for Radioastronomy in Bonn, where he
focused on studying radio-optical properties of active galactic
nuclei (AGN), kinematics of blazar jets, and on the Galactic centre physics. Currently, he is a postdoctoral researcher at the Centre for Theoretical Physics of the Polish Academy of Sciences (Prof. B. Czerny group), where his research focuses on the Broad-Line Region in AGN.}
\end{biography}

\end{document}